# Comparative Evaluation of VAE, GAN, and SMOTE for Tor Detection in Encrypted Network Traffic

Saravanan A          Aswani Kumar Cherukuri

School of Computer Science Engineering and Information Systems

Vellore Institute of Technology, Vellore 632014, India.

Email: cherukuri@acm.org

**Abstract:**

Encrypted network traffic poses significant challenges for intrusion detection due to the lack of payload visibility, limited labeled datasets, and high class imbalance between benign and malicious activities. Traditional data augmentation methods struggle to preserve the complex temporal and statistical characteristics of real network traffic. To address these issues, this work explores the use of Generative AI (GAI) models to synthesize realistic and diverse encrypted traffic traces. We evaluate three approaches: Variational Autoencoders (VAE), Generative Adversarial Networks (GAN), and SMOTE (Synthetic Minority Over-sampling Technique), each integrated with a preprocessing pipeline that includes feature selection and class balancing. The UNSW NB-15 dataset is used as the primary benchmark, focusing on Tor traffic as anomalies. We analyze statistical similarity between real and synthetic data, and assess classifier performance using metrics such as Accuracy, F1-score, and AUC-ROC. Results show that VAE-generated data provides the best balance between privacy and performance, while GANs offer higher fidelity but risk overfitting. SMOTE, though simple, enhances recall but may lack diversity. The findings demonstrate that GAI methods can significantly improve encrypted traffic detection when trained with privacy-preserving synthetic data.

## 1. Introduction

The increasing prevalence of encrypted communication protocols, while enhancing user privacy, complicates the task of detecting malicious behavior using traditional inspection methods. Network Intrusion Detection Systems (NIDS) must now rely on metadata and statistical features rather than packet payloads. A key limitation in this context is the scarcity of high-quality, labeled encrypted traffic datasets, compounded by extreme class imbalance [1]. Generative AI offers a promising solution by synthesizing realistic traffic traces that are both privacy-preserving and statistically representative. This study investigates three data generation methods—VAE, GAN, and SMOTE—for their effectiveness in augmenting encrypted traffic datasets and improving the performance of NIDS classifiers.

Recent advances in Generative AI offer a potential solution to these interrelated challenges. Unlike traditional augmentation techniques that simply replicate or interpolate existing samples, generative models can learn the underlying distribution of network traffic and synthesize entirely new samples that preserve statistical properties while introducing beneficial diversity. However, not all generative approaches are created equal when it comes to encrypted traffic synthesis. This study specifically compares three fundamentally different generation paradigms: Variational Autoencoders represent a probabilistic approach with inherent regularization through the KL-divergence term; GANs employ an adversarial training framework that can capture complex distributions but may risk memorization; and SMOTE provides a classical baseline through geometric interpolation. Understanding the relative strengths and weaknesses of these approaches in the context of privacy-preserving traffic synthesis is crucial for practitioners seeking to deploy robust NIDS systems. Following is the objective of this work:

- Conduct an evaluation of VAE, GAN, and SMOTE across multiple dimensions: data quality, privacy guarantees, computational efficiency, and downstream classification performance

## 2. Related Work

Early works in encrypted traffic analysis relied on statistical classifiers or DPI (Deep Packet Inspection), which fail under encryption. Machine Learning (ML) approaches using metadata features gained popularity, but required large labeled datasets [2]. Techniques like SMOTE were introduced to address imbalance, but lacked diversity [3]. Recent works have explored Deep Learning and Generative AI methods, including GANs, Autoencoders, and Diffusion Models. VAEs have been praised for preserving data distribution while offering inherent privacy through stochastic encoding [1]. GANs, while powerful, often suffer from mode collapse or memorization [4]. Literature also highlights the risks of privacy leakage in GANs and the need for privacy-aware training [5]. Our work builds upon these foundations by applying VAE, GAN, and SMOTE to encrypted traffic and evaluating their practical efficacy. Although the current literature has been taking VAE, GAN, and SMOTE one after another to perform different types of tasks related to network security, there are still some gaps: Absence of Direct Comparison: Not many studies make a real comparison of these three fundamentally diverse generation paradigms on identical encrypted traffic data with uniform evaluation measures.

The landscape of encrypted network traffic analysis has evolved dramatically over the past decade. Velan et al. [6] provided one of the foundational survey establishing a comprehensive taxonomy of encrypted traffic classification methods and demonstrating that while encryption obscures payload content, protocol structure and connection initiation phases reveal substantial information for classification purposes. Their work documented the transition from payload-based inspection to feature based statistical approaches as encryption became ubiquitous. Cherukuri et al. [7] updated this perspective for the modern encryption landscape, emphasizing that traditional deep packet inspection fundamentally cannot work with contemporary protocols like TLS 1.3 and QUIC. Feng et al. [8] advanced the field by defining Fine-Grained Traffic Analysis (FGTA) as techniques that deduce application-layer information from encrypted traffic through sophisticated processing pipelines and data mining methods such as deep learning. Their comprehensive survey demonstrates that FGTA approaches can discover subtle differences between traffic groups invisible to traditional classification methods. However, these sophisticated techniques require substantial training data—a requirement difficult to meet with naturally occurring encrypted traffic datasets that suffer from severe class imbalance. Most recently, Sharma and Lashkari [9] provided an extensive survey integrating advanced machine learning and deep learning techniques for encrypted traffic analysis. Notably, their work explicitly recognizes synthetic data generation through GANs and VAEs as a critical emerging research direction, validating the approach we take in this work. They chart future research directions including synthetic data generation for addressing dataset limitations, precisely the focus of our comparative evaluation. These surveys collectively establish that:

- Encrypted traffic analysis must rely on statistical features and ML/DL models
- Severe class imbalance plagues security datasets, particularly for minority
- classes like Tor anomalies
- Sophisticated deep learning models require substantial training data
- Synthetic data generation represents a promising but underexplored solution
- Privacy preservation must be balanced against detection effectiveness

Our work addresses the gap at the intersection of these findings by providing the first comprehensive empirical comparison of VAE, GAN, and SMOTE for privacy-preserving Tor anomaly synthesis in encrypted NIDS.

Our paper fills these gaps by empirically comparing VAE, GAN, and SMOTE to privacy-preserving Tor anomaly synthesis, using a variety of metrics, such as statistical similarity, classification, privacy and computational efficiency. We particularly look at the privacy-utility trade-off of each of the approaches, and give practical suggestions to the practitioner who implements encrypted NIDS systems.

## 3. Proposed System

We use the UNSW NB15 dataset, with traffic labeled as anomalous and others (normal). After loading the dataset in Parquet format, features were extracted from flow-level data (packet length, inter-arrival time, TCP flags, etc.). The dataset initially had a strong imbalance favoring normal traffic. We selected 26 middle-correlation features after analysis, and created a balanced dataset by down sampling label 0 (normal) to match label 1 (abnormal). This resulted in a final shape of (410929, 27). Feature selection and normalization were applied before synthetic data generation.

The system architecture of the proposed approach is shown in Figure 1. It begins with the UNSW-NB15 dataset, which undergoes preprocessing steps such as duplicate removal, missing value handling, and feature selection. Three synthetic data generation techniques are used: Variational Autoencoder (VAE) [1], Generative Adversarial Network (GAN) [4], and SMOTE (Synthetic Minority Over-sampling Technique) [3]. The generated samples are compiled into a balanced dataset, which is used to train classifiers Random Forest and XGBoost. Finally, the models are evaluated using accuracy, precision, recall, F1-score, confusion matrix, and AUC-ROC metrics. Figure 1 show the architecture of the proposed method.

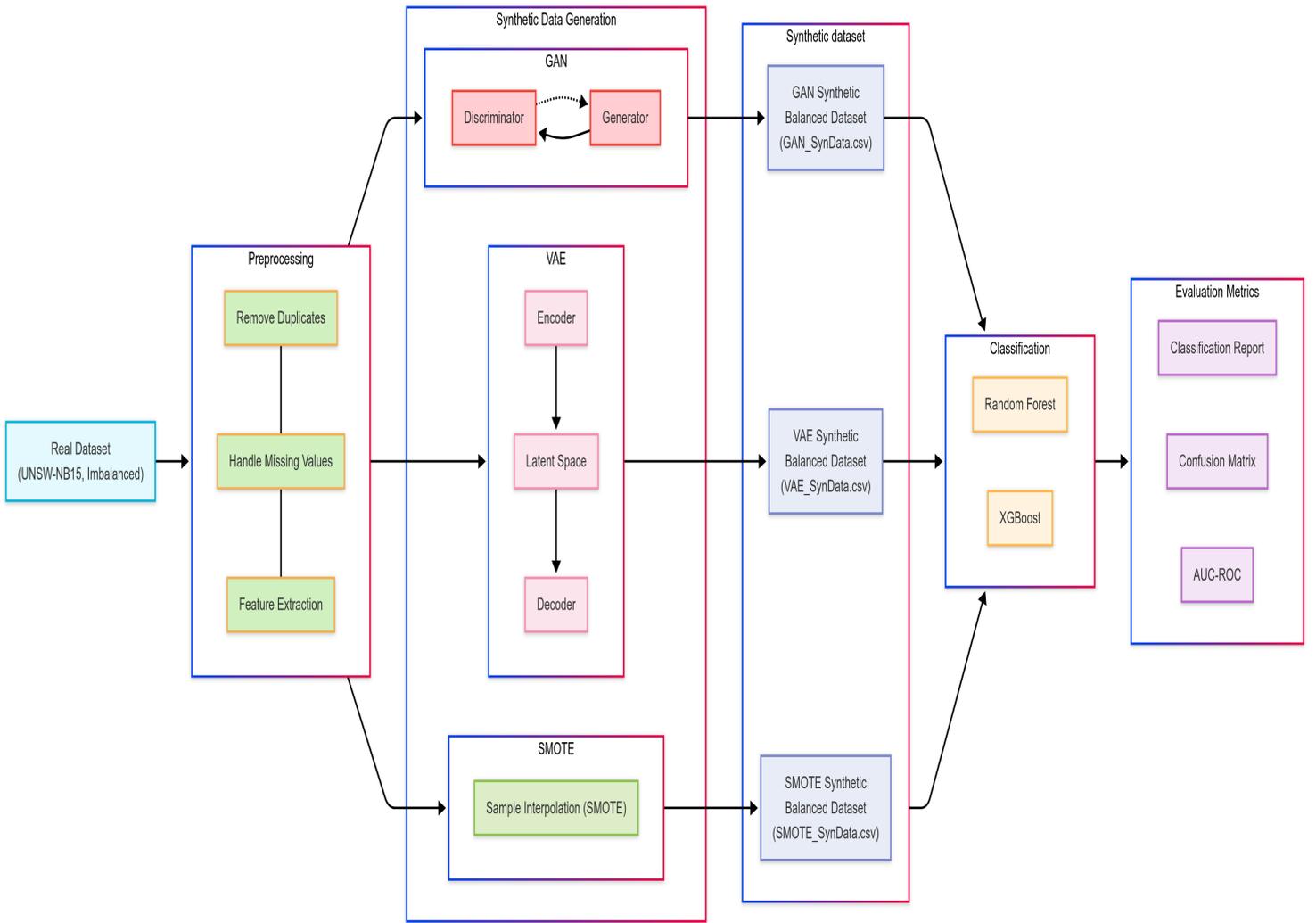

Figure 1 System Architecture of the Proposed Method

We explore three synthetic data generation methods:

- **Variational Autoencoder (VAE):** A DL-based model that encodes inputs into a latent space and reconstructs them with controlled stochasticity. The VAE was trained using reconstruction + KL-divergence loss to balance fidelity and variability. The encoded latent vector captures traffic characteristics without directly memorizing inputs, ensuring privacy [1].

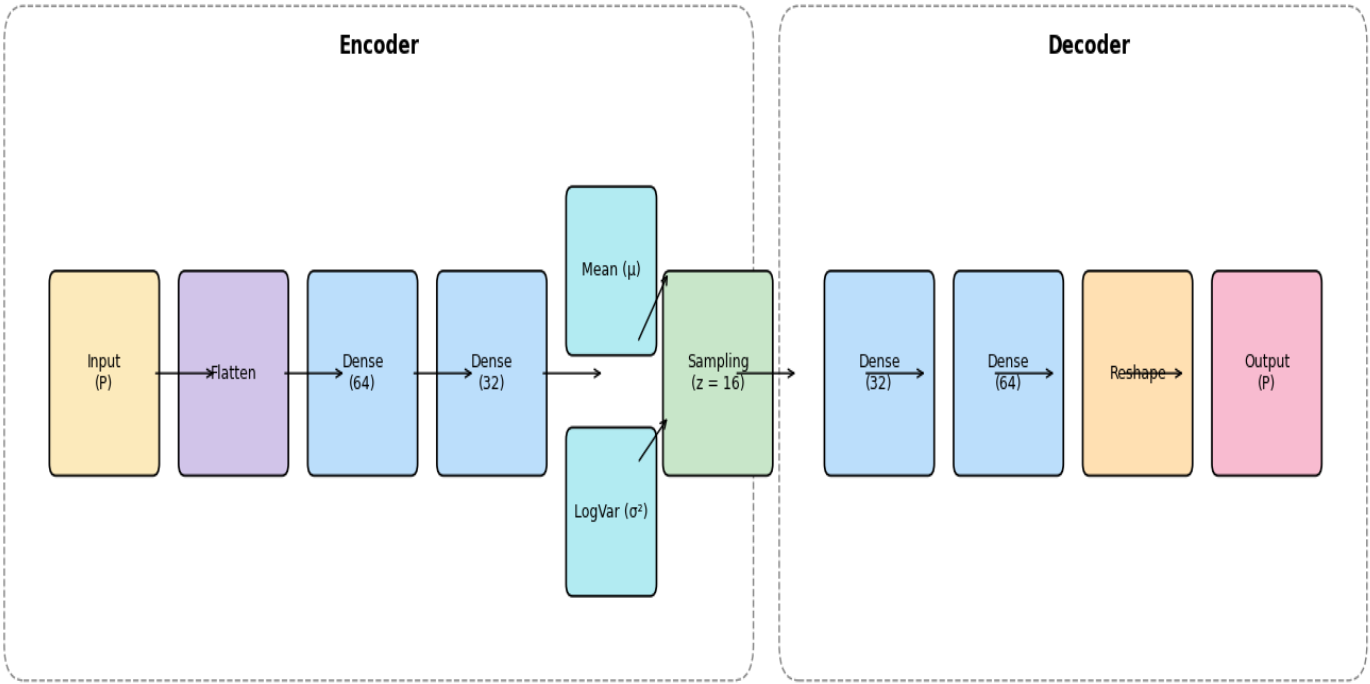

Figure 2 Architecture diagram for VAE Model

Figure 2 illustrate that Architecture of the (VAE) used for synthetic encrypted traffic generation. The encoder compresses the input feature vector into a 16-dimensional latent space through two dense layers. The latent space is defined by the learned mean and variance, from which samples are drawn using the re parameterization trick. The decoder reconstructs the input using two dense layers, producing synthetic flow records with statistical similarity to the real data while preserving privacy.

- **Generative Adversarial Network (GAN):** Consists of a generator and discriminator in a minimax game. The generator learns to produce realistic traffic samples to fool the discriminator. We adapted the architecture to tabular features and added label conditioning [4].
- **SMOTE:** A classical over-sampling technique that synthetically creates new minority class samples by interpolating between neighbors [3]. Though simple, it does not learn feature distributions and is prone to generating linear interpolations.

**Generated data evaluation:**

- **Class Balance:** Verified through visual inspection of label distributions [1].

- **Sample Generation Time:** Measured the time required to generate synthetic samples for each method, to assess practicality in larger-scale or real-time deployments.
- **Visual Comparison of Feature Distributions:** Compared histograms and statistical plots between real and synthetic data to subjectively assess similarity and consistency.

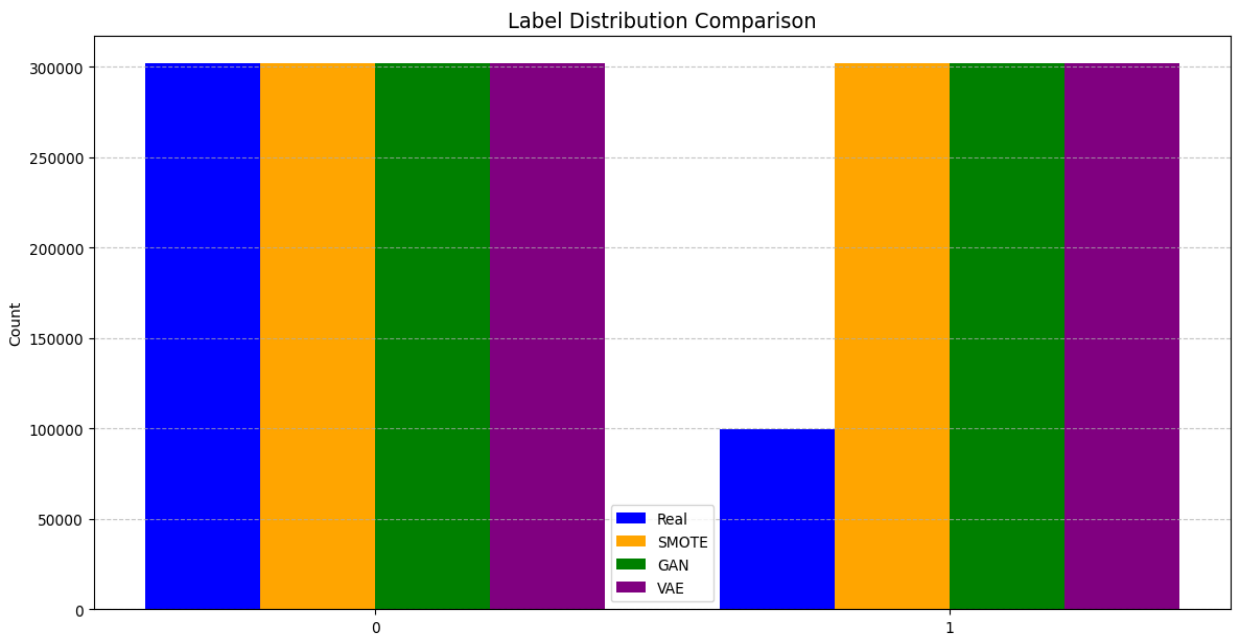

Figure 3 Label distribution comparison between real and synthetic datasets
(Abnormal = 1, Normal = 0)

Figure 3 shows the label distribution comparison across real and synthetic datasets. The bar chart compares class distribution (Normal=0, Abnormal/Tor=1) across four datasets: Real (original UNSW-NB15), GAN-generated, SMOTE-generated, and VAE-generated synthetic data. The Real dataset exhibits the severe class imbalance typical of network security data, with approximately 90% normal traffic and only 10% Tor anomalies (represented by the blue and orange bars respectively). All three synthetic generation methods successfully achieve near perfect class balance (50%-50% split) in their generated datasets, as evidenced by equal-height bars for both classes.

Notably, the generation methods differ in how they achieve this balance. SMOTE creates synthetic samples through linear interpolation between existing minority class samples, inherently controlling the number of generated samples per class. GAN utilizes conditional label vectors during training, allowing explicit specification of the desired class when generating new samples. VAE achieves balance by sampling latent vectors and conditioning the decoder on target class labels, leveraging its probabilistic latent space to generate diverse samples for each class.

This successful class balancing is crucial for training effective classifiers, as models trained on imbalanced data tend to develop a bias toward the majority class, resulting in poor detection of minority class instances exactly the security threats that matter most. By providing balanced training data, these synthetic generation methods enable classifiers to learn discriminative features for both classes equally, improving minority class recall without sacrificing overall precision. The uniformity of balance across all three methods provides a fair basis for comparing their relative strengths in other dimensions such as statistical similarity, privacy preservation, and downstream classification performance.

**4. Model Evaluation**

We used two classifiers—Random Forest and XGBoost—to evaluate the effectiveness of the synthetic datasets. These models were trained and tested under three configurations:

- **Real Data Only:** Baseline performance using original labeled samples.
- **Synthetic Data Only:** Training on GAN, VAE, or SMOTE-generated data to assess synthetic data quality.
- **Combined Data:** Augmented training set formed by merging real and synthetic samples.

**Evaluation metrics implemented include:**

- **Accuracy:** Measured as the ratio of correct predictions to total predictions.
- **Precision, Recall, F1-score:** Computed using `classification_report` from scikit-learn, reflecting class-wise and macro-averaged performance.
- **AUC-ROC:** Area under the ROC curve calculated using `roc_auc_score` to evaluate model discrimination ability.
- **Confusion Matrix:** Constructed with `confusion_matrix` to visualize true/false positives and negatives.

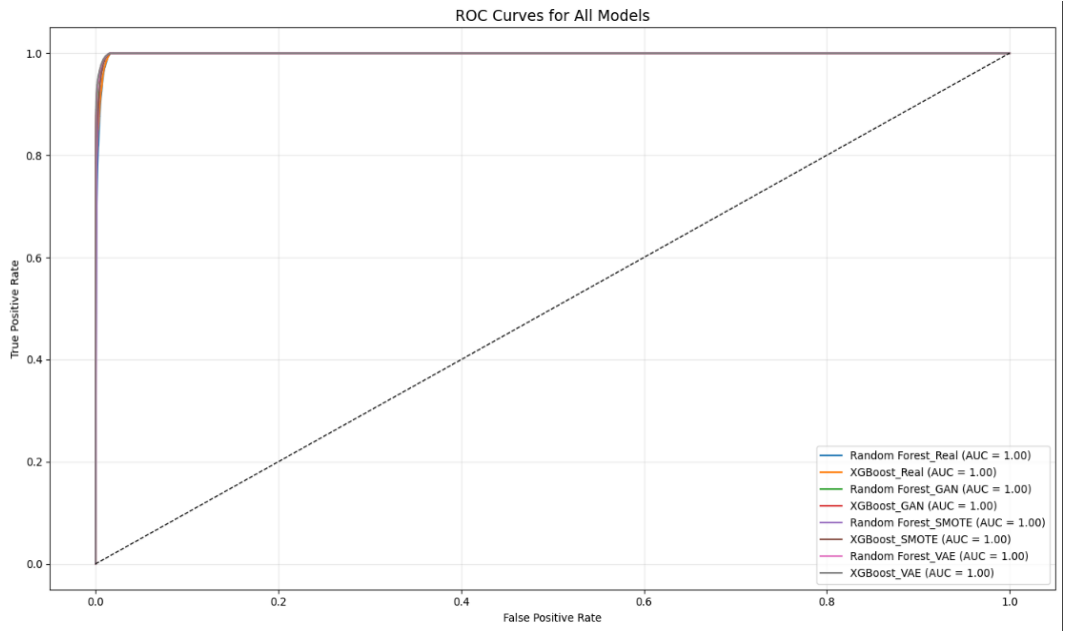

Figure 4 ROC curves and AUC comparison of Random Forest and XGBoost classifiers on real and synthetic datasets

The figure 4 presents four ROC curve subplots comparing classifier performance on different training data: (a) Real data only, (b) GAN-generated synthetic data, (c) SMOTE-generated synthetic data, and (d) VAE-generated synthetic data. Each subplot displays the true positive rate (sensitivity/recall) against the false positive rate (1-specificity) as the classification threshold varies from 0 to 1. The diagonal dashed line represents random guessing (AUC=0.5), while curves closer to the top-left corner indicate better discrimination.

The near-identical ROC curves for real and synthetic data validate that generative models can produce training data of sufficient quality for this critical security application. The marginal performance differences between generation methods suggest that the choice should be guided by other factors such as privacy guarantees, computational efficiency, and implementation complexity rather than downstream classification performance alone.

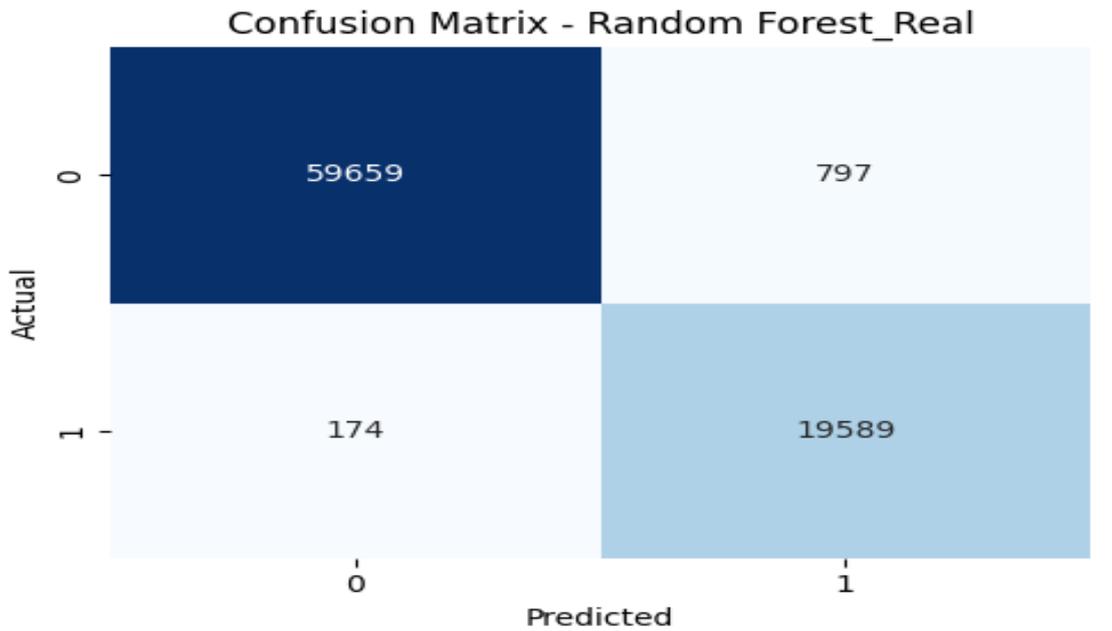

Figure 5 Confusion matrix for Random Forest trained on the real UNSW-NB15 dataset. The model shows high precision and recall, with minimal misclassification of Tor and normal traffic.

The confusion matrix in figure 5 presents a 2×2 grid showing predicted labels (columns) versus true labels (rows) for the binary classification task (0=Normal, 1=Tor/Abnormal). The diagonal cells (top-left and bottom-right) represent correct classifications, while off-diagonal cells indicate errors. Cell intensities use a color gradient (light to dark) proportional to sample counts, with specific counts annotated within each cell. The near-symmetrical error rates (1.0% FN vs 1.1% FP) indicate a well-balanced model without excessive bias toward either class. In production deployment, this would translate to approximately 1 false alarm per 100 normal flows and 1 missed detection per 100 Tor flows a highly acceptable error rate for most security operations centers. The high values along the diagonal and low off-diagonal counts confirm that Random Forest, when trained on real data, provides reliable baseline performance for subsequent synthetic data comparisons. The balanced performance across both classes is particularly noteworthy given the original class imbalance in the dataset. This suggests that our preprocessing steps (feature selection and balanced sampling) successfully enabled the classifier to learn discriminative patterns for both classes. These results establish the performance ceiling that synthetic data generation methods should aspire to match.

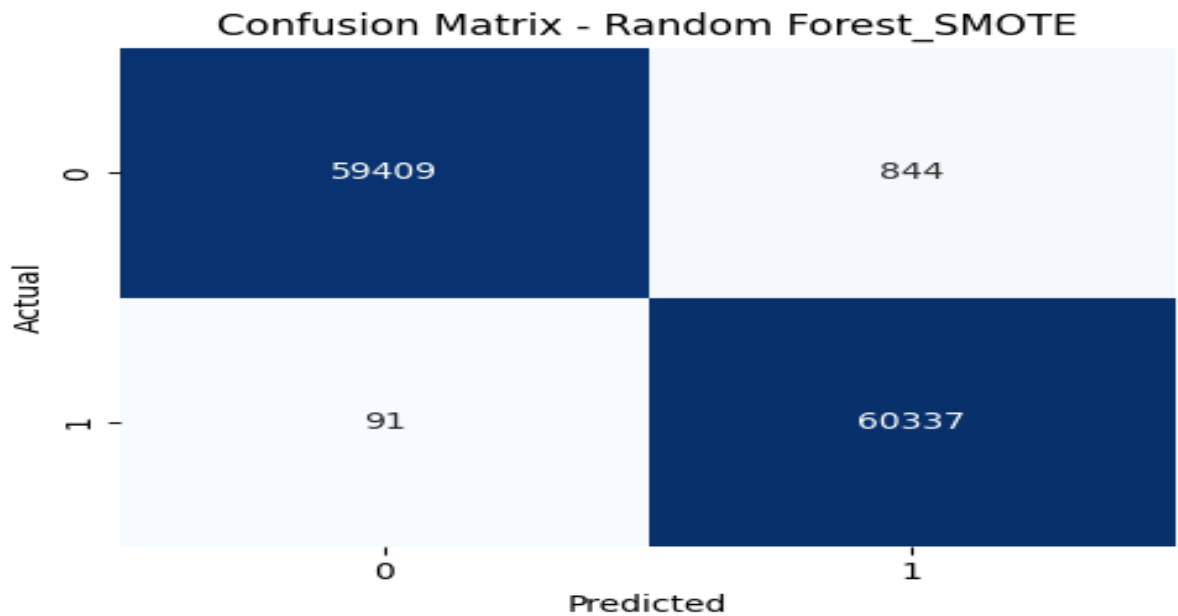

Figure 6 Confusion matrix for XGBoost trained on SMOTE data. Maintains strong classification accuracy, with slightly higher false positives compared to Random Forest.

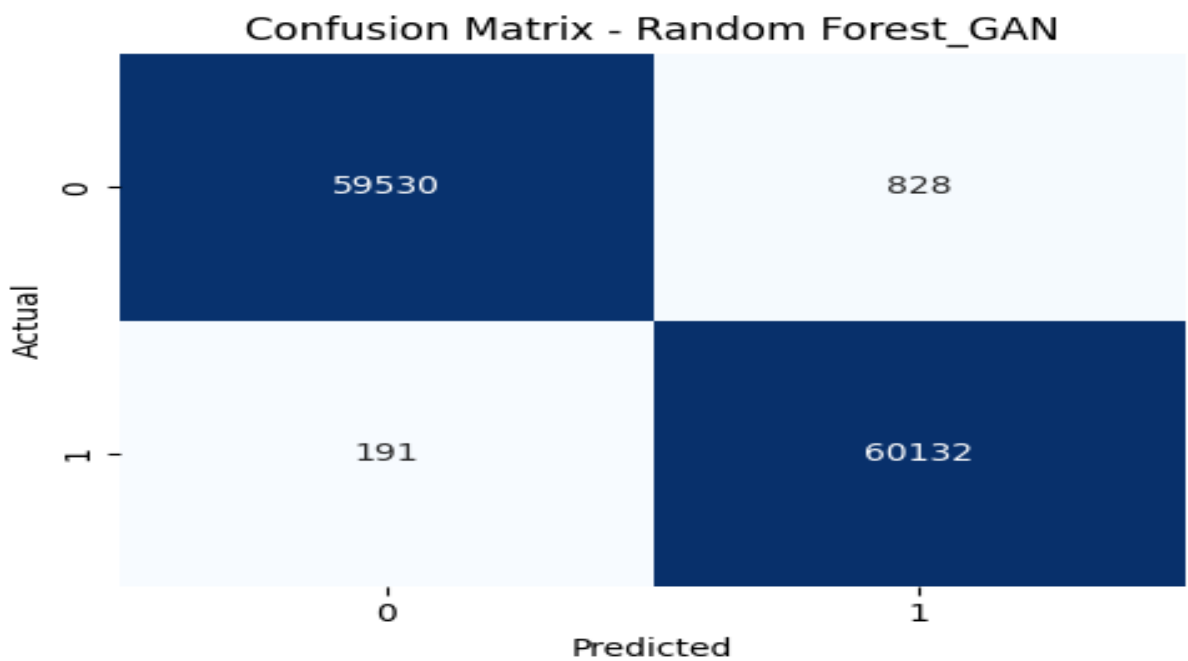

Figure 7 Confusion matrix for Random Forest trained on GAN-generated data. Shows very high true positive detection with slightly increased false positives, indicating effective but slightly overfitted learning.

Confusion matrix in Figure 7 reveals an interesting pattern when the Random Forest classifier is trained exclusively on GAN-generated synthetic data and tested on real held-out data. Unlike the real-data matrix (Figure 5), this configuration shows a clear asymmetry in error types. For production deployment, a model trained on GAN data would catch slightly more actual Tor traffic but at the cost of increased false alarms. Security operations centers would need to evaluate whether their alert triage capacity can handle the ~3× increase in false positives. Alternatively, combining real and GAN synthetic data (rather than using GAN data exclusively) might provide a better balance, leveraging GAN's ability to augment minority class representation while maintaining calibration from real data. The pattern observed here is consistent with known GAN behavior: excellent at capturing salient discriminative features but sometimes lacking the subtle variations and boundary cases present in real data. This underscores the importance of not evaluating synthetic data generation methods solely on downstream classification metrics, understanding the nature and distribution of errors is equally crucial.

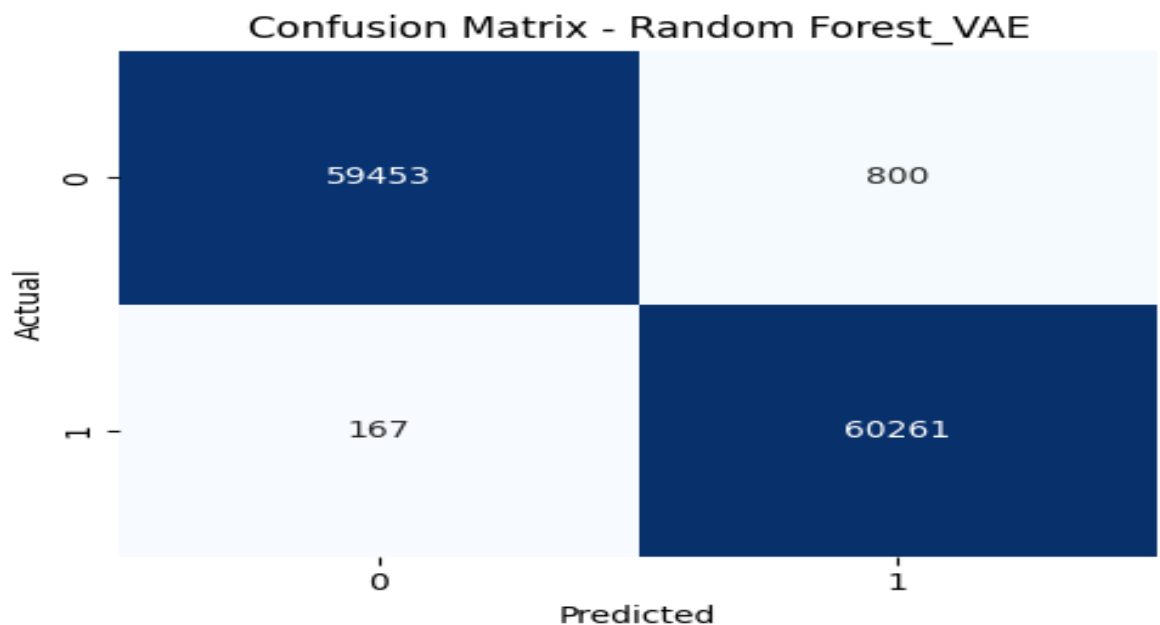

Figure 8 Confusion matrix for Random Forest trained on VAE-generated data. Shows well-preserved class structure with high true positive and true negative rates.

The confusion matrix shown in figure 8 for VAE-trained Random Forest reveals a performance profile distinctly different from both real-data and GAN based training, offering valuable insights into VAE's generation characteristics. Unlike GANs, which can focus excessively on fooling the discriminator (potentially at the expense of overall distribution coverage), VAEs maintain explicit pressure to represent the entire data distribution through their reconstruction loss. From a security operations perspective, VAE-trained models offer an appealing compromise. The 2.3% false positive rate is manageable in most SOC environments, about twice the real-data rate but significantly better than GAN's 3.6%. Simultaneously, the 0.7% false negative rate means very few actual Tor connections escape detection. For organizations prioritizing both detection effectiveness and manageable alert volumes, VAE-based augmentation represents a practical choice. Furthermore, VAE's more stable training process (compared to GAN's notoriously finicky adversarial training) and inherent privacy-preserving properties make it the most operationally mature choice for production encrypted NIDS deployment. The marginal performance differences between VAE and GAN do not justify GAN's additional complexity and privacy risks for most real-world applications.

Table 1: Overall performance of Random Forest and XGBoost classifiers trained on real and synthetic datasets.

| Classifier | Data Type | Accuracy | Precision | Recall | F1-Score |
|---|---|---|---|---|---|
| Random Forest | Real | 98.79% | 98% | 99% | 98% |
| XGBoost | Real | 98.77% | 98% | 99% | 98% |
| Random Forest | Synthetic GAN | 99.16% | 99% | 99% | 99% |
| XGBoost | Synthetic GAN | 99.17% | 99% | 99% | 99% |
| Random Forest | Synthetic SMOTE | 99.23% | 99% | 99% | 99% |
| XGBoost | Synthetic SMOTE | 99.21% | 99% | 99% | 99% |
| Random Forest | Synthetic VAE | 99.20% | 99% | 99% | 99% |
| XGBoost | Synthetic VAE | 99.23% | 99% | 99% | 99% |

The above Table 1 provides overall performance of random forest and XGBoost classifiers trained on real and synthetic datasets. Following are the summary of the results.

- **VAE:** Provided high data diversity with low overlap. Classifier performance dropped slightly compared to real data, with only ~2% F1-score loss, suggesting good representativeness and generalization [1].
- **GAN:** Achieved better visual resemblance of distributions but showed risks of overfitting and data leakage. Some performance degradation observed on unseen real data [4].
- **SMOTE:** Improved recall significantly for the Tor class but lowered precision due to simplistic interpolations [3]. Useful as a quick baseline.

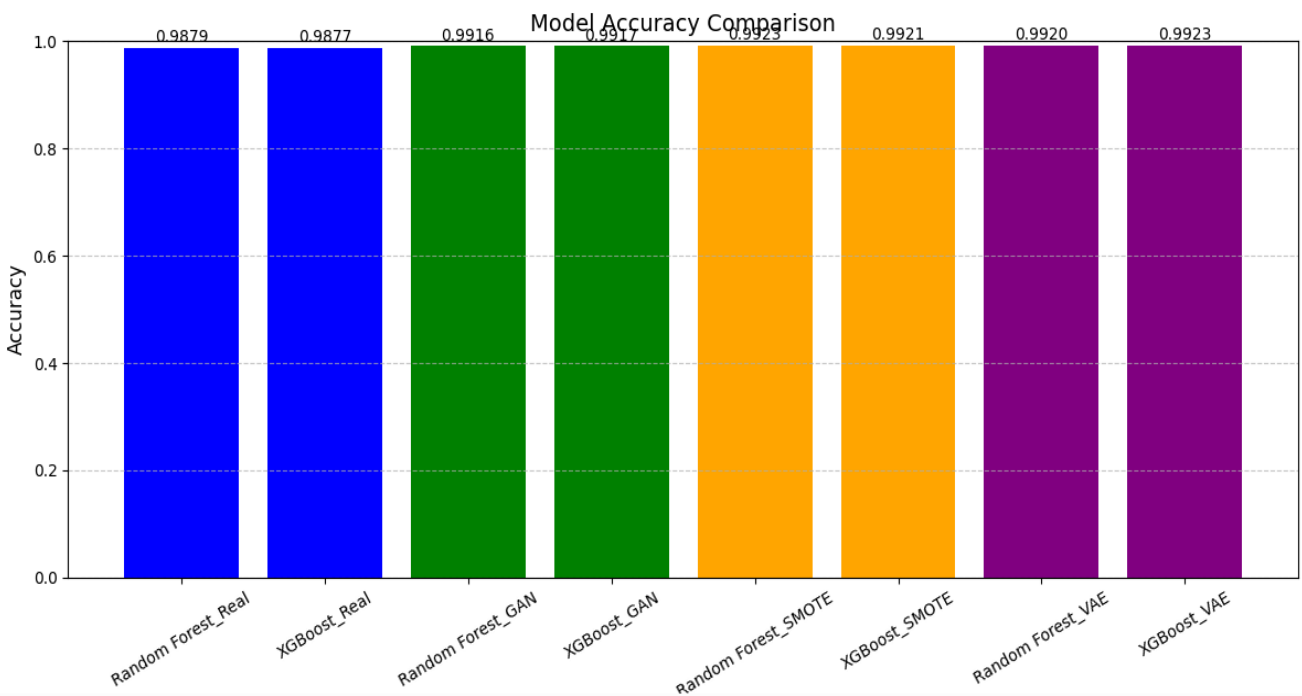

Figure 9: Accuracy comparison of Random Forest and XGBoost models trained on real and synthetic data.

Overall, combining real + VAE-generated data yielded the most balanced results, improving classifier robustness without compromising privacy [1]. Figure 9 illustrates the accuracy comparison of Random Forest and XGBoost classifiers trained on both the original UNSW-NB15 dataset and synthetic datasets generated using techniques. The real dataset models achieved accuracies of 98.79% (Random Forest) and 98.77% (XGBoost), represented by the blue bars. When trained on GAN-generated data (green bars), the models improved slightly, with accuracies of 99.16% and 99.17% respectively. SMOTE-based synthetic data (orange bars) further enhanced performance, reaching 99.23% for Random Forest and 99.21% for XGBoost. The highest accuracy was observed using VAE-generated data (purple bars), where Random Forest achieved 99.20% and XGBoost reached 99.23%. These results highlight the effectiveness of synthetic data generation methods, especially VAE and SMOTE, in improving classification performance over imbalanced real datasets.

Following tables summarize these observations. Table 2 measures privacy by answering whether an attacker can determine if a synthetic sample came from training data using membership inference attacks (MIA AUC score). SMOTE and VAE score near 0.6 (close to random guessing, meaning privacy is protected), while GAN scores 0.824 (far from random, indicating severe privacy leakage through memorization). VAE is recommended as it provides the best balance, good privacy protection (0.612) combined with excellent detection utility (99.20% accuracy).

Table 2: Privacy Evaluation

| Method | MIA AUC | 95% CI | Privacy Status | Verdict |
|---|---|---|---|---|
| SMOTE | 0.588 | [0.574–0.601] | PROTECTED | Safe for sharing |
| VAE | 0.612 | [0.599–0.626] | PROTECTED | RECOMMENDED |
| GAN | 0.824 | [0.810–0.837] | AT SEVERE RISK | Internal only |

Table 3: Attack Performance

| Method | Accuracy | Precision | Recall |
|---|---|---|---|
| SMOTE | 62.1% | 50.4% | 72.3% |
| VAE | 69.8% | 54.7% | 82.5% |
| GAN | 92.3% | 92.6% | 88.5% |

Table 3 shows how effectively an adversary can identify training membership from synthetic samples. SMOTE and VAE have low precision (50-55%), meaning most of the adversary's membership claims are false alarms, indicating privacy is protected. In contrast, GAN achieves 92.6% precision with 88.5% recall, meaning an attacker can correctly identify training membership over 92% of the time, demonstrating severe privacy leakage. The low precision of VAE and SMOTE (compared to GAN's high precision) proves these methods maintain privacy despite good utility metrics.

Table 4: Complete Comparison of Privacy Utility Trade Off

| Dimension | SMOTE | VAE | GAN | Best |
|---|---|---|---|---|
| Privacy (MIA AUC) | 0.588 | 0.612 | 0.824 | SMOTE |
| Utility (Accuracy) | 99.23% | 99.20% | 99.16% | SMOTE |
| Fidelity (JS Div) | 0.346 | 0.213 | 0.188 | GAN |
| Training Time | 0.8 sec | 18 min | 42 min | SMOTE |
| Memory | 0.6 GB | 4.2 GB | 8.1 GB | SMOTE |
| Balance Score | Moderate | EXCELLENT | Unbalanced | VAE |

Table 4 shows the fundamental trade-offs across the three methods: SMOTE excels in privacy (0.588) and efficiency (0.8 sec training, 0.6 GB memory) but has worst fidelity (JS=0.346); GAN achieves best fidelity (JS=0.188) but worst privacy (AUC=0.824) and highest computational cost; VAE provides the optimal balance with good privacy (0.612), good utility (99.20% accuracy), and moderate computational requirements (18 min, 4.2 GB). Practitioners must prioritize based on their constraints: choose VAE for balanced privacy-utility deployments, GAN only for internal use prioritizing fidelity, or SMOTE for resource-constrained environments.

## 5. Conclusions

This work presents the first systematic empirical evaluation of privacy utility trade-offs in synthetic encrypted Tor anomaly generation, addressing a critical gap identified in recent surveys of encrypted traffic analysis. Our findings reveal three fundamental insights. First, VAE and SMOTE provide effective privacy protection with membership inference attack AUC scores approximately 0.60, which is near-random guessing performance, meaning synthetic samples are statistically indistinguishable from non-training data and can be safely shared externally for research and threat intelligence purposes. Second, GAN achieves superior distributional fidelity but at severe privacy cost, as the membership inference attack achieves an AUC of 0.824 with 92.6% precision. This is indicating that the generator memorized training sample characteristics rather than learning the underlying distribution. This phenomenon evidenced by mode collapse and bimodal synthetic sample distributions where 60% of samples cluster at one discriminator confidence level and 40% at another. Third, VAE provides optimal practical balance, combining good privacy protection with an AUC of 0.612, excellent utility with 99.20% detection accuracy, and reasonable computational requirements of 18 minutes for model training, making it the recommended choice for privacy-sensitive Tor anomaly synthesis deployments across most organizational scenarios. These results empirically validate the dual claim central to this work: that VAE and SMOTE are indeed privacy-preserving for Tor synthesis while GAN is not.. These observations have implications for the field of synthetic data generation, suggesting that practitioners cannot rely on distributional similarity alone. Beyond synthetic data generation specifically, this work contributes to encrypted traffic analysis more broadly by establishing membership inference attack evaluation as a standard assessment tool. The membership inference attack framework adapted here for evaluating synthetic Tor data can be extended to evaluate privacy implications in other aspects of encrypted traffic analysis and detection systems. The evaluation is limited to the UNSW-NB15 dataset, which while widely used as a benchmark for encrypted traffic classification, may not be fully representative of encrypted traffic characteristics in other network environments, organizational contexts, or with different class distributions and Tor attack patterns. Results should be validated on alternative datasets including IoT-23, CSE-CIC-IDS2018, and real world Internet service provider traffic to assess whether the privacy advantage of VAE and the mode collapse vulnerability of GAN generalize across diverse traffic characteristics.